\documentclass[twoside,reqno]{HERON}
\usepackage{epsfig,cite,colordvi}
\usepackage{graphicx}
\usepackage{url}
\usepackage{amssymb,amsmath,amscd,epsf}
\usepackage{times}
\usepackage{makeidx}
\usepackage{amssymb,amsmath,bm,mathtools}
\usepackage{siunitx,booktabs}
\usepackage{graphics,graphicx,epstopdf,epsfig}
\usepackage{tabularx}
\usepackage{multirow}
\usepackage{psfrag}
\usepackage{array}
\usepackage{hyperref}
\usepackage[english]{babel}

\usepackage{adjustbox}

\def\nbR{\ensuremath{\mathrm{I\! R}}}

\newcommand{\be}{\begin{equation}}
\newcommand{\ee}{\end{equation}}
\newcommand{\ba}{\begin{eqnarray}}
\newcommand{\ea}{\end{eqnarray}}

\begin{document}

\title{Determination of the moments of the proton charge density: is there a proton radius puzzle?
}

\runningheads{Determination of the moments of the proton charge density: is there a proton radius puzzle?
}{M. Atoui, M.B. Barbaro, M. Hoballah, C. Keyrouz, R. Kunne, M. Lassaut, D. Marchand, G. Qu\'em\'ener, E. Voutier
}

\begin{start}

 \coauthor{M. Atoui}{1}, \author{M.B. Barbaro}{2,3}, \coauthor{M. Hoballah}{1}, \coauthor{C. Keyrouz}{1,4}, \coauthor{R. Kunne}{1}, \coauthor{M. Lassaut}{1}, \coauthor{D. Marchand}{1}, \coauthor{G. Qu\'em\'ener}{5}, \coauthor{E. Voutier}{1}, \coauthor{J. van de Wiele}{1}

\index{Barbaro, M.B.}
\index{Atoui, M.}
\index{Hoballah, M.}
\index{Keyrouz, C.}
\index{Kunne, R.}
\index{Lassaut, M.}
\index{Marchand, D.}
\index{Qu\'em\'ener, G.}
\index{Voutier, E.}
\index{van de Wiele,  J.}

\address{IJCLab, Orsay, F-91405 Orsay, France}{1}

\address{Universit\`a di Torino, Dipartimento di Fisica, I-10125 Turin, Italy}{2}

\address{Istituto Nazionale di Fisica Nucleare, Sezione di Torino, I-10125 Turin, Italy}{3}

\address{INPHYNI, Nice, F-06560 Valbonne, France}{4}

\address{LPC, Caen, F-14050 Caen, France}{5}

\begin{Abstract}
The charge radius of the proton can be determined using two different kinds of experiments: the spectroscopy technique, measuring the hyperfine structure of hydrogen atoms, and the scattering technique,  deducing the radius from 
elastic lepton scattering off a proton target. These two methods lead to quite different results, a discrepancy  known as the "proton radius puzzle ".
To shed light on this problem, we have proposed a novel method for the determination of spatial moments from densities expressed in the momentum space. This method provides a direct access not only to the second order moment, directly related to the proton radius, but to all moments of any real order larger than -3. 
The method is applied to the global analysis of proton electric form factor experimental data from Rosenbluth separation and low-$Q^2$ experiments, paying specific attention to the evaluation of the systematic errors. Within this analysis, the integer order moments of the proton charge density are evaluated, the moment of second order leading to a new determination of the proton charge radius.
\end{Abstract}
\end{start}

\section{Introduction}

The determination of the proton radius has been the object  of intense scientific activity   in the last decade on both theoretical and experimental sides (see Refs.~\cite{Car15,Hil17,Kar20,Gao22}  for comprehensive reviews). 
The  proton charge  radius, defined as~\cite{Atoui:2023nrn}
\begin{equation}
R_p \equiv \sqrt{ -6 \left. \frac{\mathrm{d} G_E(k^2)}{\mathrm{d}k^2} \right\vert_{k^2=0} } \,,
\label{redef} 
\end{equation}
 is related to the slope of the proton electric form factor $G_E$ at vanishing squared four-momentum $k^2$.
Two experimental techniques have been proposed and developed  to measure $R_p$: the spectroscopy technique, which provides a measurement of the  proton radius from the hyperfine structure of ordinary or muonic hydrogen atoms,  and the scattering technique, where the value of $R_p$ is deduced from the cross section of elastic lepton scattering off a proton target. 

The so-called "proton radius puzzle"~\cite{Ber14} originated from the substantial discrepancy between the electron scattering measurement~\cite{Ber10} and the muonic spectroscopy one~\cite{Poh10}, providing the values 0.879(8)~fm  and  0.84184(67)~fm, respectively. This discrepancy has raised criticisms on the scattering method, suggesting that the extrapolation procedure of experimental data to zero-momentum transfer suffers from limited accuracy: since lepton scattering cannot reach the zero four-momentum transfer limit, this technique relies on the zero-momentum extrapolation of $G_E(k^2)$ and then strongly depends on the functional form as well as on the data analysis method used for the extrapolation~\cite{Lee15,Sic17}. Consequently, the scattering technique is 
intrinsically less accurate than the spectroscopy one.
This appears to be a blatant limitation of the scattering technique, particularly difficult to overcome, as also indicated by the discrepancies between the latest scattering measurements of the proton radius~\cite{Ber10, Xio19, Mih17}. It  also  
hampers any attempt to determine higher-order moments of the proton charge density through this approach - here-after referred to as the derivative method. 
Additionally, moments of the charge density beyond the second order are also of interest as they carry complementary information on the charge distribution inside the proton. However, beyond the limited precision of the experimental determination of higher derivatives of the form factor, the derivative method accesses only even moments of the density.
The recent PRad result~\cite{Xio19} and the recommended CODATA~\cite{COD20} and PDG~\cite{Zyl20} values of $R_p$ have reduced the tension with muonic atom measurements; nevertheless, improving the precision of scattering experiments remains a high priority in light of the numerous discussions about the sensitivity of the derivative method (see Ref.~\cite{Bar2020} for new developments).

With the aim of overcoming the limitations inherent  to  the derivative method in the scattering technique as well as of improving the predictive power  of the technique,  we have proposed a novel approach~\cite{Hob20}, referred to in the following as the integral method, that enables the determination of spatial moments of densities for any real-valued order $\lambda$$>$-$3$. 
  In Sec.2  this method will be described and  its validation illustrated in a specific case. In Sec.3 its application to real experimental data for the proton electric form factor, as performed in Ref.~\cite{Atoui:2023nrn}, will be presented. In Sec.4 some conclusions will be drawn and future possible developments will be outlined.

%
%
\section{The integral method}
\label{sec:spatmom}

Before introducing the integral method (IM) we briefly review the basic formalism of the standard derivative method.
The  moments   $\langle r^\lambda \rangle$ of the charge density $\rho_E({\bold r})$ are defined as
\be
 \langle r^\lambda \rangle
\equiv \int_{\nbR^3}  \mathrm{d}^3\bold{r} \, r^\lambda \rho_E({\bold r})\, .
\label{eq:rlam}
\ee
Within a non-relativistic description of the internal structure of the proton they can be related, for the case of even order $\lambda=2j$, to the derivatives of the electric form factor $G_E(k^2)$ at $k^2=0$
\begin{equation} \langle r^{2j} \rangle = (-1)^j \, \frac{(2j+1)!}{j!} \, \left. \frac{\mathrm{d}^j G_E(k^2)}{\mathrm{d} (k^2)^j} \right\vert_{k^2=0} \label{eq:there} \end{equation}
if the relation
\begin{equation}
\rho_E({\bold r}) = \frac{1}{(2\pi)^3} \int_{\nbR^3} \mathrm{d}^3\bold{k} \, e^{ i \bold{k}\cdot \bold{r}} G_E(k^2) \, ,
\label{eq:TF}
\end{equation}
which links $\rho_E(\bold r)$ to its Fourier transform $G_E(k^2)$, is used. Note that here $k^2=\bold k^2$,  a relation holding true in the Breit frame.
In particular, for $j=1$ one recovers  Eq.~\eqref{redef} for the radius 
$R_p = \sqrt{ \langle r^2 \rangle }$.
Therefore, using the derivative method one can evaluate all the moments of even and positive order by extrapolating the derivatives of the form factor $G_E$ to $k^2=0$. This procedure suffers from the uncertainties described in the Introduction.

Alternatively, one can start from the general expression 
\begin{equation} \label{eq:IM0}
\langle r^{\lambda} \rangle = \frac{1}{(2\pi)^3}\ \int_{\nbR^3} \mathrm{d}^3\bold{k} \, G_E(k^2) \int_{\nbR^3}  \mathrm{d}^3\bold{r} \, e^{i \bold{k}\cdot \bold{r}} r^{\lambda} 
\end{equation}
for the moment \eqref{eq:rlam} of  any order $\lambda$, which is obtained  exploiting Eqs.~\eqref{eq:rlam} and \eqref{eq:TF}  and interchanging the integrals  over $\bold{k}$ and  $\bold{r}$.
The IM consists in evaluating the radial moments using the expression \eqref{eq:IM0}, after some appropriate manipulation which will be detailed in the following.
While the derivative method crucially depends on the way the form factor approaches its value at zero momentum, the integral method involves the full $k^2$ physics region. This conceptual change is of importance for the determination of the functional form of $G_E(k^2)$, which can be expected to be more precise, and likewise for the corresponding moments.
Moreover, unlike  the derivative method, Eq.\eqref{eq:IM0} allows one to evaluate moments of  any real order,  the only limitation being the convergence of the integral.

To further develop the expression \eqref{eq:IM0}, we observe that the left-hand side of 
the  above equation, the moment $\langle r^{\lambda} \rangle$, is a finite quantity which represents a physics observable; however, the right-hand side  contains the integral  
\be \label{sect2eq03}
g_{\lambda}(\bold{k}) 
= \int_{\nbR^3} \mathrm{d}^3\bold{r} \, e^{i \, \bold{k}\cdot \bold{r} } r^{\lambda} \, .
\ee
This integral does not exist in a strict sense, but it can still be treated as a distribution; the finiteness of the left-hand side ensures the physical  representativity of this expression as well as the convergence of the 6-fold integral. For instance, for $\lambda=0$  Eq.~\eqref{sect2eq03} corresponds to the Dirac distribution $\delta(\bold  k)$ which, inserted in \eqref{eq:IM0}, provides 
\be
\langle r^0 \rangle \equiv \int_{\nbR^3}  \mathrm{d}^3\bold{r} \,  \rho_E({\bold r}) = G_E(0)\,.
\ee
One then needs to regularize the function $g_{\lambda}(\bold{k})$, using standard methods. In Ref.~\cite{Hob20} two different regularization schemes were presented, one based on the subtraction of counter-terms and the other on the introduction of an exponential function inside the integral \eqref{sect2eq03}, which makes the integral convergent and is taken equal to 1 at the end of the calculation. Although the two methods lead to  identical numerical results, it  has been proven in \cite{Hob20} that  the latter provides a faster numerical  convergence. Here we briefly describe only this method, which will be used in the application described in Sec.3.
Within this approach, the distribution $g_{\lambda}(\bold{k})$ is expressed as a  limit of a convergent integral, namely
\begin{eqnarray}\label{sect2eq03p}
g_{\lambda}(\bold{k}) = \lim_{\epsilon \to 0^+} \int_{\nbR^3} \mathrm{d}^3 \bold{r} \, r ^{\lambda} e^{-\epsilon r} \, e^{i \, \bold{k}\cdot \bold{r} } = \lim_{\epsilon \to 0^+} {\mathcal I}_{\lambda}(k,\epsilon)\,,
\end{eqnarray}
where the term $e^{-\epsilon r}$ ensures the convergence of the integral ${\mathcal I}_{\lambda}(k,\epsilon)$ and ``$\lim$" should be understood as a weak limit~\footnote{that is, the limit should be taken after  performing the integral over $\bold k$.}. This is a standard technique used, for example, to regularize the Fourier transform of the Coulomb potential~\cite{Fet80,Alt19}. The integration of Eq.~\eqref{sect2eq03p} is analytical and yields, for any $\lambda > -3$ and $\lambda \neq -2$,
\begin{equation} \label{Ilambda}
{\mathcal I}_{\lambda}(k,\epsilon) = \frac{4 \pi \, \Gamma(\lambda+2) \, \sin\left[ (\lambda+2) {\rm Arctan} \left( k/\epsilon \right) \right]}{k (k^2 + \epsilon^2)^{\frac{\lambda}{2}+1} } \,,
\end{equation}
which attains the limit $(4 \pi/k) {\rm Arctan} \left( k/\epsilon \right)$ at $\lambda$=$-2$. The moments defined in Eq.~\eqref{eq:rlam} can then be written as
\begin{eqnarray}\label{sect2eq05p}
\langle r^{\lambda} \rangle =  \frac{2}{\pi}  \, \Gamma(\lambda+2) 
 \lim_{\epsilon \to 0^+} \int_{0}^{\infty} \mathrm{d}k \, G_E(k^2) \, \frac{k \sin\left[ (\lambda+2) {\rm Arctan} \left( k/\epsilon \right) \right]}{(k^2 + \epsilon^2)^{\lambda/2+1}} \,.
\end{eqnarray}
 For integer orders, $\lambda=m$, Eq.~\eqref{sect2eq05p} can also be recast as
\begin{eqnarray}
\label{eq:rm}
\langle r^m\rangle  =  \frac{2}{\pi} \, (m+1)!  \label{sect2eq08} \lim_{\epsilon \to 0^+} \epsilon^{m+2} \int_0^{\infty} \mathrm{d}k \, G_E(k^2) \, \frac{k \, \Phi_{m}(k/\epsilon) }{ (k^2 + \epsilon^2)^{m+2} }
\end{eqnarray}
with
\begin{equation} \label{eqfim}
\Phi_{m}(k/\epsilon) =\sum_{j=0}^{m+2} \sin \left( \frac{j \pi}{2} \right) \frac{(m+2)!}{j!(m+2-j)!} \, \left( \frac{k}{\epsilon} \right)^j \, .
\end{equation}
This formulation  allows us to determine the moments directly in momentum space, for both integer and non-integer values of $\lambda$. For a given $G_E(k^2)$ functional form, the moments are numerically computed from the above expressions and can also be obtained analytically for specific cases. 

As a simple illustration, let us consider  the radial density
\begin{equation} \label{dip-den}
\rho_D  ( \bold{r} ) = \frac{\Lambda^3}{8\pi} \, e^{-\Lambda r} \,,
\end{equation}
leading to the well-known dipole parameterization
\begin{equation}\label{ffdip}
{G}_E^{(D)}(k^2) = \int_{\nbR^3} \mathrm{d}^3\bold{r} \, e^{- i \bold{k}\cdot \bold{r}} \rho_D (\bold{r}) = \frac{\Lambda^4}{(k^2+\Lambda^2)^2} \,,
\end{equation}
where $\Lambda$ represents the dipole mass parameter. In this case the moments can be easily determined directly in configuration space, as 
\begin{equation} \label{dip-mom}
\langle r^{\lambda}\rangle_D = \int_{\nbR^3} \mathrm{d}^3\bold{r} \, r^{\lambda} \, f_D(\bold{r}) = \frac{\Gamma(\lambda+3)}{2} \, \frac{1}{\Lambda^{\lambda}} \, .
\end{equation}
For integer $\lambda$=$m$ values, Eq.~\eqref{eq:rm} yields (see \cite{Hob20} for the detailed calculation)
\begin{equation}
\langle r^{m}\rangle_D = \frac{2 \, \Gamma(m+2)}{\pi} \, \frac{1}{\Lambda^m} \, \lim_{\tilde{\epsilon} \to 0^+} \frac{\pi}{4} \frac{m+2}{(1+\tilde{\epsilon})^3} = \frac{\Gamma(m+3)}{2} \frac{1}{\Lambda^{m}} \,, \label{rmfD}
\end{equation}
which coincides with the result \eqref{dip-mom} obtained from the configuration space integral. The same result is obtained for any real (integer and non-integer) $\lambda$ value from the numerical evaluation of the integral \eqref{sect2eq05p}. 

The IM has also been  tested for different mathematical realizations of the radial function $\rho(\bold r)$ and several $\lambda$, in particular for a Yukawa-like  density  corresponding to the parameterization of the proton electromagnetic form factors in terms of a $k^2$-polynomial ratio, the Kelly's parameterization of Eq.~\cite{Kel04}
\begin{equation}
\label{kellymodel}
G_E(k^2) = 
\frac{1+a_1 k^2}{1+b_1 k^2 + b_2 k^4 + b_3 k^6}\,.
\end{equation}
Also in this case, the numerical evaluation of the integral \eqref{sect2eq05p} provides with a very high accuracy the same results as the configuration space integrals.

As already mentioned, the IM overcomes the limitation of the derivative method to moments of positive even-valued orders. This is a remarkable advantage, since each moment order of the charge density is of interest, as it carries complementary information on the charge distribution inside the proton. For instance, the short-distance behaviour of the charge distribution is encoded in the negative order moments which are particularly sensitive to the large $k^2$-dependence of $G_E$, while the long-distance behaviour is encoded in the high positive order moments.
However, the evaluation of moments via this method requires an experimentally defined asymptotic limit which may be hardly obtained considering the  momentum coverage of actual experimental data. The momentum dependence of the integrand in  Eq.~\eqref{sect2eq05p} provides the solution to this issue. The denominator of the integrand scales at large momentum like $k^{\lambda+1}$, meaning that the integral is most likely to saturate at a momentum value well below infinity. 

Truncated moments, defined from Eq.~\eqref{sect2eq08} and Eq.~\eqref{sect2eq05p} by replacing the infinite integral boundary by a cut-off $Q$, allow us to understand the saturation behaviour of the moments. Considering for sake of simplicity the case of integer $\lambda$=$m$ values, the truncated moments can be written from Eq.~\eqref{sect2eq08} as
\be
\langle r^m\rangle_Q = \frac{2}{\pi} \, (m+1)! \,  \lim_{\epsilon\to 0^+} \epsilon^{m+2} \, \int_0^Q \mathrm{d}k \, G_E(k^2)\, \frac{k \, \Phi_m(k/\epsilon)}{(k^2+\epsilon^2)^{m+2}} \,.
 \label{Rinte}\ee
For the typical example of the dipole parameterization of Eq.~\eqref{ffdip}, the truncated moments of even order are given by
\be
\langle r^{2p}\rangle_{Q} 
= \frac{(2p+2)!}{2} \, \frac{1}{\Lambda^{2p}} \label{eq:eve0}
\ee
and are independent of $Q$, while the odd-order  ones
\be
\langle r^{2p+1}\rangle_{Q} = \frac{2}{\pi} \, {(2p+2)!} \,  \left[ u_{2p+1}(Q,0^+) + v_{2p+1}(0^+) \, {\rm Arctan} \left(\frac{Q}{\Lambda}\right) \right]  \label{eq:odd0}
\ee
are still depending on the cut-off. Indeed, Eq.~\eqref{eq:eve0} can be seen as a proof of the $Q$-independence of even moments. The explicit expressions for the functions $u_i$ and $v_i$ can be found in Ref.~\cite{Hob20}. The $u_i$'s functions  behave like $1/Q$ at large cut-off, and consequently vanish for infinite $Q$; only the $v_i$'s remain in the infinite-$Q$ limit, leading to the expression of Eq.~\eqref{rmfD}. 

\begin{figure}[h]
\centering
\includegraphics[width=0.5\columnwidth]{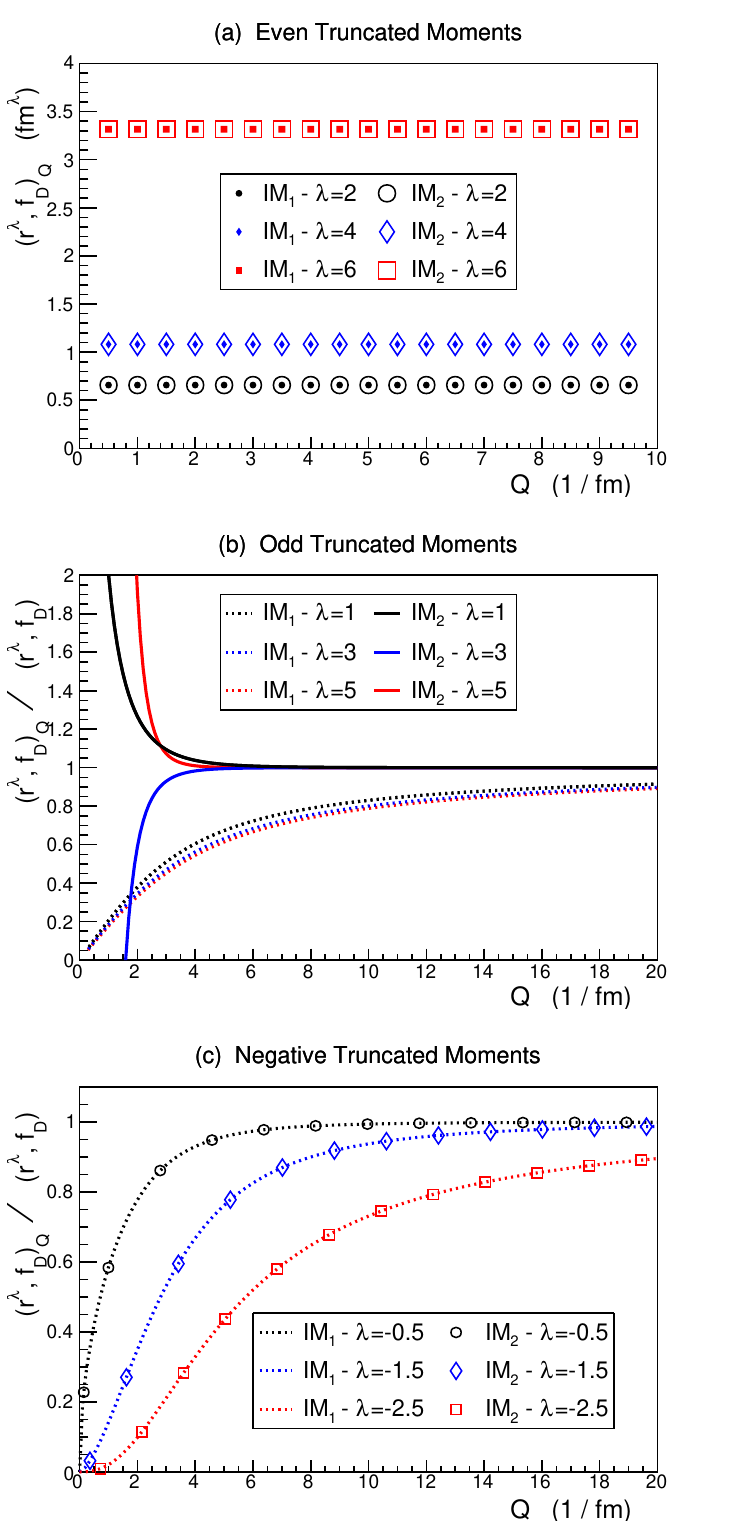}
\caption{Convergence of truncated moments of the proton electric form factor for selected orders within the dipole parameterization: (a) positive even, (b) positive odd, and (c) negative non-integer. IM$_1$ and IM$_2$ denote the principal value and the exponential regularizations, respectively.}
\label{fig:Qsat} 
\end{figure}

The $Q$-convergence of truncated moments is illustrated in Fig.~\ref{fig:Qsat} in the case of the dipole form factor for selected moment orders, as obtained for the two previously mentioned regularization  schemes, based on counterterms subtraction (IM$_1$) or on the introduction of an exponential convergence factor (IM$_2$).
The $Q$-independence of even truncated moments is reproduced by both prescriptions, as shown in Fig.~\ref{fig:Qsat}(a). This is a general feature independent of the specific form factor:  for even moments the integral method recovers formally the same quantities as the derivative method. In the ideal world of perfect experiments, adjusting experimental data with the same function over a small or large $k^2$-domain affects only the precision on the parameters of the function. In the context of the limited quality of real data, the IM provides the mathematical support required to consider the full $k^2$-unlimited domain of existing data, leading therefore to a more accurate determination of the moments. The practical constraint is to obtain an appropriate description of the data over a large $k^2$-domain.  
 Fig.~\ref{fig:Qsat}(b) shows the $Q$-convergence of selected odd-order moments, comparing the integral method prescriptions. In this case the different regularization schemes  lead  to different saturation behaviours: while IM$_1$ asks for large $Q$-values, IM$_2$ rapidly saturates about $6$~fm$^{-1}$, {\it i.e.} in a momentum region well covered by proton electromagnetic form factors data~\cite{Pun15}. 
 Fig.~\ref{fig:Qsat}(c) shows the $Q$-convergence of selected moments with negative non-integer orders. For such orders, there are no counterterms for the IM$_1$ regularization  and the effect of the exponential regularization term in Eq.~\eqref{sect2eq03p} is strongly suppressed since the integrand converges at infinity (for $-3<\lambda<-1$). Indeed, there is no need of regularization for negative orders and all prescriptions should lead to identical results: this is verified in Fig.~\ref{fig:Qsat}(c) where the numerical evaluation of each prescription is shown to provide the same result for $-3<\lambda<0$. 

Similar features can be derived  for the polynomial ratio parameterization \eqref{kellymodel}, 
leading  to the the truncated moments~\cite{Hob20}
\ba
&& \langle r^{-2}\rangle_{_Q} = \frac{1}{2} \sum_{n=1}^3 A_n \ln{\left( 1 + \frac{Q^2}{(ik_n)^2} \right)} 
\nonumber 
 \\
 && \langle r^{2p-1}\rangle_{_Q} = (2p)! \frac{2}{\pi} \, \sum_{n=1}^3 \frac{A_n}{(ik_n)^{2p+1}} \left[ {\mathrm{Arctan}{\left( \frac{Q}{ik_n} \right)} - \frac{Q}{ik_n}  } 
- \sum_{j=0}^{p} \frac{(-1)^j}{2j-1} {\left( \frac{ik_n}{Q} \right)}^{2j-1} \right] 
 \nonumber\\ && 
\langle r^{2p}\rangle_{_Q} = (2p+1)! \sum_{n=1}^3  \frac{A_n}{(ik_n)^{2p+2}} \, , 
\label{eq:tr3}
\ea
where $ik_n$ are the  poles of the form factor \eqref{kellymodel} in the positive imaginary plane and $A_n$ the corresponding residues.

Summarizing,  for practical applications the entire $k^2$ physics region can be restricted as the high $k^2$-region may not contribute significantly to the integral. This is particularly true for positive order moments which, considering the $k^2$-dependence of experimental data, are dominated by the region $k^2 \le 2$~GeV$^2$~\cite{Hob20}. The situation is more tricky for negative order moments due to their sensitivity to large $k^2$.

 %
%
\section{Application to experimental data} \label{sec:app}

Following the formal demonstration of the integral method, we now present  the experimental evaluation of a selected set of $\lambda$-order moments of the proton charge density from the proton electric form factor data.

The data inputs consist of proton electric form factor data $G_E(k^2)$ extracted from electron scattering experiments via a Rosenbluth separation~\cite{Ros50} or for kinematical conditions where the  contribution of the magnetic form factor $G_M(k^2)$ to the cross section is strongly suppressed, for instance at very low four momentum transfer. 
The experimental data set considered consists of 21 
single data sets  covering  $k^2$ values up to 226~fm$^{-2}$ (8.8~GeV$^2$) (see Ref.~\cite{Atoui:2023nrn} for the list of experiments and  the  selection critera). 

\begin{figure}[ht!]
\centering
\includegraphics[width=0.5\columnwidth]{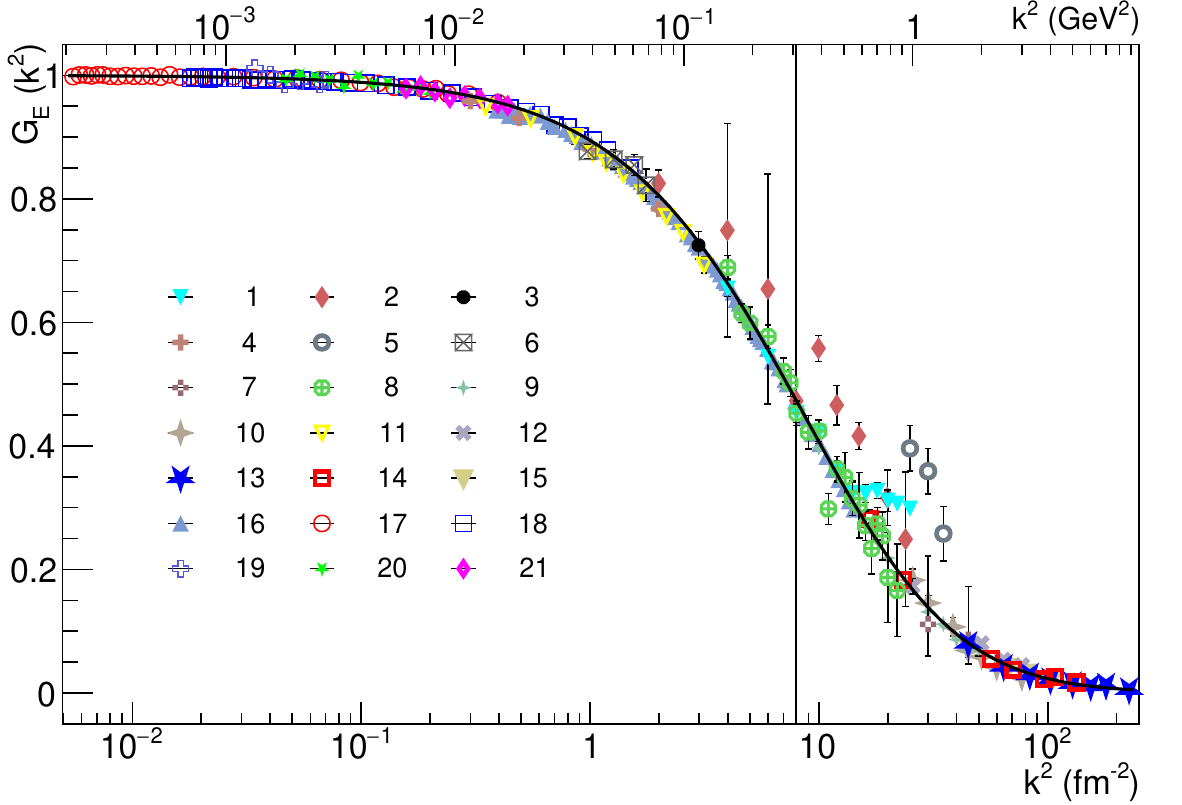}%
\includegraphics[width=0.5\columnwidth]{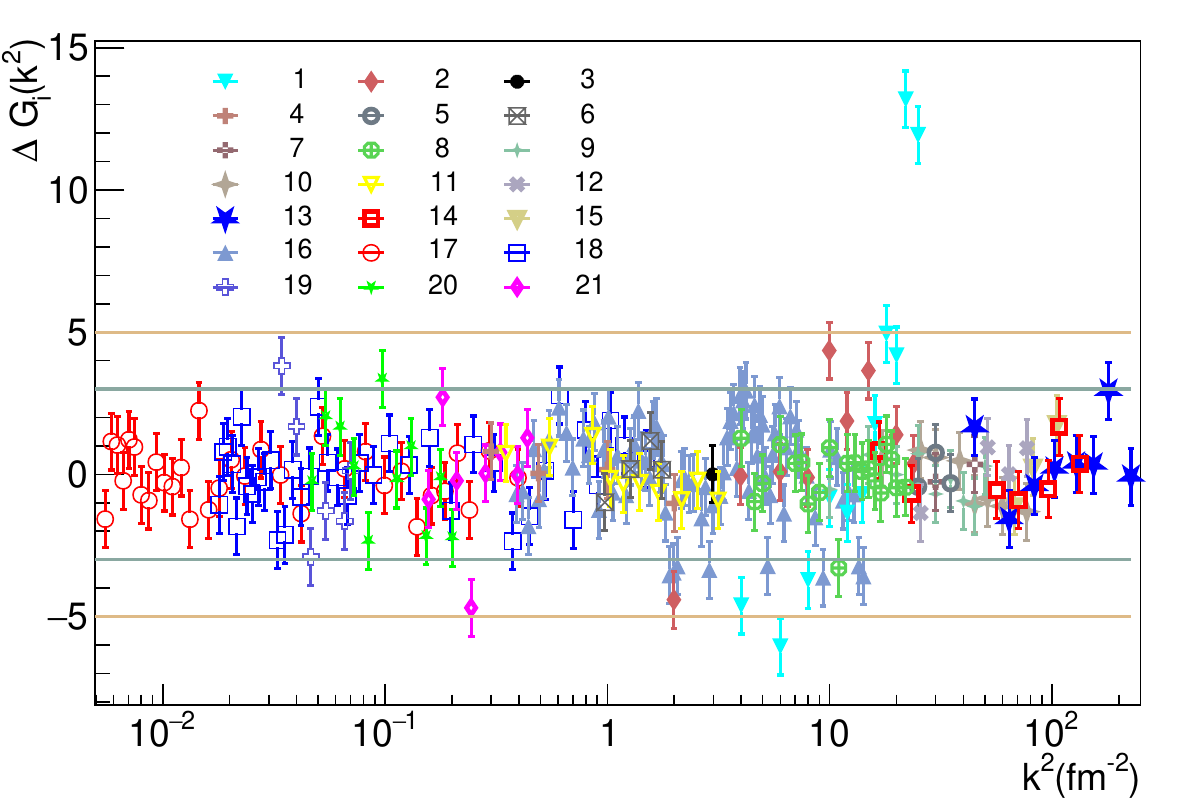}%
\caption{Left: Simultaneous fit (black line) of $G_E(k^2)$ experimental data conside\-red in this work 
using the polynomial ratio function of Eq.~\eqref{kellymodel}. Right: residual deviation of the experimental data with respect to the fit.
}
\label{FFit} 
\end{figure}

The complete set is analyzed within a simultaneous fit approach requiring the same $k^2$-dependence for each experiment and a separate normalization factor per data set. The chosen fitting function is in the polynomial ratio \eqref{kellymodel} 
multiplied by a normalization parameter $\eta_i$ for each data set $i$. This method follows the analysis techniques of the most recent experiments~\cite{Ber14-1,Xio19,Mih21}. The results of the best fit to experimental data are represented in Fig.~\ref{FFit}, where the 
residual deviations $\Delta G_i(k^2)$ of each experimental data set ($i$) from the fit is also reported. 
The fit accounts for a reduced $\chi^2_r$=$2.96$, quite reasonable considering the actual data dispersion.

The experimental moments of the proton charge density are determined from the form factor functional using the integral method restricted to the measured physics region. 
In the absence of experimental data at very large four momentum transfer, pertubative QCD provides scaling rules which predict a rapid decrease of $G_E(k^2)$~\cite{Bro73}. Thus, the effect of the truncation of the integral 
can be controlled. The integration cut-off $Q^2$=52~fm$^{-2}$ considered in the  present study was proven to only impact the evaluation of negative order moments~\cite{Hob20}. 

The values of the truncated moments, obtained according to Eqs.~\eqref{eq:tr3}, are reported in Tab.~\ref{Momres} for odd and even orders in the range $-2 \le \lambda \le 7$, together with their asymptotic limits 
for $Q \to \infty$. Note that the 0$^{\mathrm{th}}$-order moment is $1$ by definition of the fit function, thus is not an experimentally determined moment. As expected, negative moments are the only ones to significantly suffer from the truncation of the moment integral. Positive even moments are also compared to values ${\langle r^{2p} \rangle}_{d}$ obtained from the derivative \eqref{eq:there}.

\begin{table}[t!]
\begin{center}
\begin{adjustbox}{max width=\columnwidth}
\begin{tabular}{c||c|c|c||c|c||c|c|c|c}
\hline
\multirow{3}{*}{$\lambda$} & \multirow{2}{*}{$\langle r^{\lambda} \rangle_{_Q}$} & \multirow{2}{*}{$\langle r^{\lambda} \rangle$} & \multirow{2}{*}{${\langle r^{2p} \rangle}_{d}$} & \multicolumn{2}{c||}{Statistical Error} & \multicolumn{4}{c}{Systematic Error} \\
 & & & & $\delta \left[ {\langle r^{\lambda} \rangle_{_Q}} \right]$ & $\delta \left[ {\langle r^{2p} \rangle_{d}} \right]$ & Dat. & Int. & Fun. & Mod.\\  
 & (fm$^{\lambda}$) & (fm$^{\lambda}$) & (fm$^{\lambda}$) & (fm$^{\lambda}$) & (fm$^{\lambda}$) & (fm$^{\lambda}$) & (fm$^{\lambda}$) & (fm$^{\lambda}$) & (fm$^{\lambda}$) \\ \hline\hline
          -2 & \phantom{-}6.5826 & \phantom{-}8.9093 &               $-$ & 0.0039 & $-$ & 0.0141 &      \phantom{-}2.3267 & 0.0008 & 0.0183 \\
          -1 & \phantom{-}1.9752 & \phantom{-}2.1043 &               $-$ & 0.0005 & $-$ & 0.0024 &      \phantom{-}0.1291 & 0.0002 & 0.0022 \\
\phantom{-}1 & \phantom{-}0.7186 & \phantom{-}0.7158 &               $-$ & 0.0004 & $-$ & 0.0025 & 0.0030 & 0.0001 & 0.0008 \\
\phantom{-}2 & \phantom{-}0.6824 & \phantom{-}0.6824 & \phantom{-}0.6824 & 0.0020 & 0.0021 & 0.0113 & $0$ & 0.0001 & 0.0053 \\
\phantom{-}3 & \phantom{-}0.7966 & \phantom{-}0.7970 &               $-$ & 0.0096& $-$ & 0.0500 & \phantom{-}0.0004 & 0.0005 & 0.0300 \\
\phantom{-}4 & \phantom{-}1.0208 & \phantom{-}1.0208 & \phantom{-}1.0208 & 0.0515 & 0.0472 & 0.2498 & $0$ & 0.0042 & 0.1752 \\
\phantom{-}5 & \phantom{-}0.9219 & \phantom{-}0.9217 &               $-$ & 0.3098 & $-$ & 1.4388 & 0.0002 & 0.0273 & 1.0995 \\
\phantom{-}6 &           -3.6823 &           -3.6823 &           -3.6823 & 2.0835 & 2.2561 & 9.5186 & $0$ & 0.2372 & 7.5914 \\
\phantom{-}7 &           -49.6804 &           -49.6802 &               $-$ & 15.8544 & $-$ & 71.5437 & 0.0002 & 1.7403 & 58.1979 \\ \hline
\end{tabular}
\end{adjustbox}
\end{center}
\caption
{Moments of the proton charge density as determined from the integral method and the derivative one for even moments; the truncated moments of the second column are evaluated for the cut-off $Q^2$=52~fm$^{-2}$; the infinite moments of the third column are similarly evaluated in the limit $Q \to \infty$, assuming that the large $k^2$-dependence of the form factor is described by Eq.~\eqref{kellymodel}; refer to the text for the determination of the statistic and systematic errors of the moments.}
\label{Momres}
\end{table}

The statistical errors of the experimental moments reported in Tab.~\ref{Momres}  are determined from the propagation of the statistical errors of the fit parameters taking into account their correlations.
The magnitude of the statistical error limits the significance of the moments  determination to $\lambda<6$. This is a consequence of the existing data set which lacks measurements at ultra low momentum transfer. High order positive moments indeed probe the long distance behaviour of the charge density and are therefore specifically sensitive to this region. High accuracy measurements in this region are extremely challenging. 

Concerning systematic errors, four different sources are considered: the one related to the systematic error of the form factor measurements, the one related to the determination method of the moments, the one intrinsic to the fit function, and a last one attached to the choice of fit function. More details on the evaluation of these errors can be found in Ref.~\cite{Atoui:2023nrn}.
The only experimental source of systematics corresponds to the error on the moment originating from the systematic errors of the measurements. 
The other systematics are specifically attached to the integral method, {\it i.e.} they do not have an experimental origin per se. Except negative moments where truncation effects are dominant, the model choice turns out to be the most significant contribution to systematic errors of the IM. It is worth noticing that this error is sometimes omitted, particularly in some of the many analyses of experimental data looking for the charge radius of the proton after the highlighting of the proton radius puzzle~\cite{Kar20}. Using a mathematical function obtained from a physics model is the only way to minimize this error. 
%
%

As a particular case, the proton charge radius is determined directly from the second moment of the charge density as
\begin{equation}
R_p = \sqrt{\langle r^{2} \rangle} = 0.8261\pm 0.0012 \pm 0.0076 \, {\rm fm}\,,
\end{equation}
 where the first error is statistical and the second is systematic.

%
%
\section{Conclusions} \label{sec:concl}

We  have proposed a new method to determine the spatial moments of densities expressed in momentum space, {\it i.e.} form factors. The method provides a direct access to real moments, both positive and negative, for any form factor functional. Particularly, it represents the only opportunity to access spatial moments when the Fourier transform of a parameterization cannot be performed. In addition, unlike the derivative method which is restricted to even moments, the  integral method gives access to moments of any order $\lambda$, especially odd $\lambda$ values and more generally any real order $\lambda > -3$. Furthermore, it provides the formal support to take into account the full range of existing data for the determination of even moments, allowing us to improve the accuracy as compared to the derivative method.

The integral method involves the regularization of integrals treated as distributions. Two regularization techniques have been tested with respect to different parameterizations of the electromagnetic form factor of the proton. In particular, it has been demonstrated that the exponential regularization provides the most performant approach, allowing us to determine accurately positive moments considering a  saturation squared four-momentum transfer of 2~GeV$^2$. Negative moments require larger saturation momenta but remain quite accessible with reduced accuracy (a few percents) in the proton case. 
 
Based on a comprehensive analysis of the proton electric form factor data obtained from Rosenbluth separation and low $k^2$ measurements, the method has been applied to the experimental evaluation of the moments of the proton charge density, paying specific attention to the determination of statistical and systematic errors. The actual status of experimental data allows a meaningful determination of the moments up to the fifth order. The sensitivity to the specific mathematical expression of the form factor fit function and the actual experiment systematics are found to be the most significant contributions to the systematic error. 
Taking into account the fit function sensitivity appears to reconcile the different determinations of the proton charge radius. In that respect, the integral method approach yields the proton charge radius value 0.8261$\pm$ 0.0012 (stat.) $\pm$ 0.0076 (syst.) fm. The current analysis suggests that the accuracy of this result, already in fair agreement with the muonic hydrogen spectroscopy value, would be improved by enriching the electric form factor data set at very low $k^2$ and using form factor functions supported by physics models.

The present study paves the way to possible future developments, including the application of the integral method to the study of the magnetic charge distribution in the nucleon. Furthermore, the generalization of Eq.~\eqref{sect2eq05p} to a $D$-dimensional  density, performed in Ref.~\cite{Hob20}, offers the  possibility to address the relativistic nature of the nucleon structure for $D$=2~\cite{Mil19}.

%
%
\section*{Acknowledgements}

This work was supported by the LabEx Physique des 2 Infinis et des Origines (ANR-10-LABX-0038) in the framework {$\ll$~Investissements d'Avenir~$\gg$} (ANR-11-IDEX-01), the French Ile-de-France region within the SESAME framework, the INFN under the Project Iniziativa Specifica MANYBODY, and the University of Turin under the Project BARM-RILO. This project has received funding from the European Unions's Horizon 2020 research and innovation programme under grant agreement No 824093.

\end{document}